\documentstyle[11pt,aaspp4,flushrt]{article}

\begin{document}

\title{Element Diffusion During Cosmological Structure Formation}

\author{David Medvigy \& Abraham Loeb}

\affil{Harvard-Smithsonian Center for Astrophysics, 60 Garden St.,
Cambridge, MA 02138; dmedvigy@cfa.harvard.edu, aloeb@cfa.harvard.edu}

\begin{abstract}

We consider element diffusion during the linear growth of structure in the
intergalactic medium (IGM) before the redshift of reionization.  The
elements produced during big bang nucleosynthesis (such as D, $^4$He or
$^7$Li) condensed in collapsing regions more readily than hydrogen.
Diffusion was most effective when the IGM was neutral since the collisional
cross--section between neutral species is smaller than the corresponding
Coulomb cross--section in an ionized gas. We find that diffusion led to
small deviations in element abundances of up to $0.1\%$.  Abundance
measurements to this level of precision in low--metallicity environments
could provide information about the thermal and ionization history of the
primordial gas during the early formation of structure in the IGM.

\end{abstract}


\section{Introduction}

Recent attempts to measure the primordial $^4$He abundance report
accuracies better than $\sim 1\%$ (Izotov \& Thuan 1998; Olive, Steigman,
\& Skillman 1997; Peimbert, Peimbert, \& Ruiz 2000).  Such accurate
measurements are useful for testing the consistency of big bang
nucleosynthesis and also for estimating the mean baryon density, $\Omega_b$
(Schramm \& Turner 1998).  The primordial $^4$He mass fraction, $\bar{Y}$,
is often inferred from the emission lines of blue compact galaxies (e.g.,
Peimbert \& Torres-Peimbert 1974; Pagel, Terlevich, \& Melnick 1986) under
the implicit assumption that it had the same initial value throughout the
Universe, including within underdense regions of the intergalactic medium
(IGM).

However, a net difference between the element abundances in overdense and
underdense regions could arise if diffusion occurs during the gravitational
collapse of massive objects.  Different atomic species, because of their
different masses, will have different thermal speeds and will therefore
experience different pressure forces per unit mass.  Consequently,
different elements would not flow into an overdense region at the same
rate; instead, heavier elements would obtain higher relative abundances in
galaxies than in voids.  Diffusion is generally moderated by a frictional
force arising from inter-species collisions.  These collisions are most
frequent for an ionized gas (Spitzer 1956) in which the characteristic
cross--section for ion-proton collisions, $\sim \pi (Ze^2/k_BT)^2 \ln
\Lambda_{\rm C} \sim 2\times 10^{-12} Z (T/10^4~{\rm K})^{-2}~{\rm cm^2}$,
is far greater than the collision cross--section for the corresponding
atoms in a neutral gas, $\sim 10^{-15}~{\rm cm^{2}}$ (here, $Ze$ is the ion
charge, $T$ is the temperature, $k_B$ is Boltzmann's constant, and~ $\ln
\Lambda_{\rm C}\sim 20$~ is the Coulomb Logarithm).  We therefore focus our
attention on the period between cosmological recombination and reionization
during which the IGM was neutral (Loeb \& Barkana 2001). During this stage,
atomic collision cross sections may be small enough to allow some diffusion
to occur.

In this paper we quantify the diffusion of $^4$He, $^7$Li, and D relative
to $^1$H that occurs as a spherical, slightly overdense region begins to
collapse.  We assume a $\Lambda$CDM cosmology, characterized by densities
relative to critical of $\Omega_m=0.3$, $\Omega_{\Lambda}=1-\Omega_m$, and
$\Omega_b=0.045$ for matter, vacuum, and baryons; a Hubble constant of
$H_0=70\ {\rm km}\ {\rm s}^{-1}\ {\rm Mpc}^{-1}$; a $^4$He primordial mass
fraction of $\bar{Y}=0.244$; and primordial $^7$Li and D number densities
relative to $^1$H of $\bar{X}_L=1.75\times 10^{-10}$ and $\bar{X}_D
=2.0\times 10^{-5}$, respectively.  The effects of diffusion are most
important where the gas pressure matters, namely around the comoving Jeans
length scale $\lambda_J=(2\pi/k_J)$.  For redshifts $z\lesssim 140$, when
the baryonic temperature is primarily determined by the adiabatic
expansion, the Jeans wavenumber is (Barkana \& Loeb 2001)
\begin{equation} 
k_J= \left[\frac{0.9 \left( 1+z \right) \mu m_H \Omega_m H_0^2}
{k_B T \left( z \right)}\right]^{1/2} 
\label{eq:k_J}
\end{equation}
where $h=({H_0}/{100\ {\rm km\ s^{-1}\ Mpc^{-1}}})$, $\mu$ is the mean
molecular weight, $m_H$ is the mass of hydrogen and $T\left( z\right) $ the
baryonic temperature at a redshift $z$.

In \S~\ref{sec:baseq}, we derive the differential equations used for our
analysis.  These are linearized equations which are strictly valid only as
long as the overdensities remain small.  However, we assume that
reionization occurs before the collapsing region becomes significantly
nonlinear and we stop our calculation at reionization because charged
particle cross sections suppress any further diffusion.  In
\S~\ref{sec:inicond}, we discuss the initial conditions used for our
differential equations.  In \S~\ref{sec:calc}, we compute both the mean
abundance deviation and the typical abundance fluctuation within a
collapsing region. We then use a Press-Schechter mass function to average
the resulting deviations over all collapsed objects in the Universe.  In
\S~\ref{sec:obs}, we compare our model to recent measurements of $^4$He
abundances.  Finally, we present our conclusions in \S~\ref{sec:sandc}.

\section{Basic Equations}
\label{sec:baseq}

We first establish our notation and illustrate our assumptions by carrying
out a brief derivation of the equations satisfied by density perturbations.
Each particle species has a mass density, $\rho_s$, and a bulk velocity,
${\bf v}_s$, which are related by the continuity equation,
\begin{equation}
\frac{\partial \rho_s}{\partial t}
+ \nabla_p \cdot \left( \rho_s {\bf v}_s \right) = 0,
\label{eq:conteq}
\end{equation}
where we start by using physical (proper) coordinates.  Since the Universe
is initially homogeneous, $\rho_s = \bar{\rho}_s \left( t \right) $ and all
particles move with the Hubble flow: ${\bf v}_s = \bar{H}\left( t \right)
{\bf x} $ (Peebles 1993).  In this approximation, equation
(\ref{eq:conteq}) becomes
\begin{equation} 
\frac{\partial \bar{\rho}_s}{\partial t} + 3 \bar{\rho}_s
\bar{H} = 0  . 
\label{eq:conteqz}
\end{equation}
We next allow for density and velocity perturbations through the 
definitions 
\begin{equation}
\rho_s \left( {\bf x},t \right) = \bar{\rho}_s\left( t\right)
\left[ 1 + \delta_s \left( {\bf x},t \right) \right] ,
\label{eq:defrhop}
\end{equation}
\begin{equation}
\partial^i v_s^j \left( {\bf x}, t \right)= 
\sigma_s^{ij}\left( {\bf x},t\right) + 
\omega_s^{ij} \left( {\bf x},t\right) 
+ \left[ \bar{H}\left( t \right) + \delta H_s \left( {\bf x},t\right) 
\right] \delta_{kronecker}^{ij} .
\label{eq:defvelp}
\end{equation}
The velocity gradient is decomposed into a symmetric, traceless tensor
($\sigma_s^{ij}$), an antisymmetric tensor ($\omega_s^{ij}$), and a trace.
We insert these definitions into equation (\ref{eq:conteq}) and linearize
the perturbations to obtain
\begin{equation}
\frac{\partial \delta_s}{\partial t} + \left( {\bf v}_s \cdot \nabla
\right) \delta_s + 3 \delta H_s =0 .
\label{eq:ceqphysco}
\end{equation}
Switching from now on from physical to comoving coordinates, and letting an
overdot represent a time derivative, equation (\ref{eq:ceqphysco}) may be
compactly expressed as
\begin{equation}
\dot{\delta_s} = -3 \delta H_s .
\label{eq:ceqcoco}
\end{equation}

The density and bulk velocity also satisfy a force equation; in comoving
coordinates, 
\begin{equation}
\dot{{\bf v}}_s = \rho_s^{-1}\sum_{forces} {\bf F}_s . 
\label{eq:forceeq}
\end{equation}  
The sum on the right-hand side includes all forces acting on species $s$.
We linearize this equation and make use of equation (\ref{eq:ceqcoco})
to obtain a new equation which does not explicitly contain the
velocity perturbations,
\begin{equation}
\ddot{\delta_s} + 2 \bar{H} \dot{\delta}_s - 3\frac{\partial \bar{H}}
{\partial t} - 3\bar{H}^2 = -\nabla\cdot \left[ \frac{1}{\rho_s}
\sum_{forces} {\bf F}_s \right]  . 
\label{eq:diffeqt}
\end{equation}

We now switch independent variables from time to redshift via the Friedmann
equation, which leads to
\begin{equation}
H_0 t = \frac{2}{3\sqrt{\Omega_{\Lambda}}}\sinh^{-1}\left(
\sqrt{\frac{\Omega_{\Lambda}}{\Omega_m}}\left( 1+z \right)^{-3/2}
\right) .
\label{eq:ztrel}
\end{equation}
However $\Omega_{\Lambda}=0.7$ only becomes significant for $z\lesssim 1$;
in this paper we consider redshifts $z\gg 1$ for which
\begin{equation}
H_0 t \approx \frac{2}{3\sqrt{\Omega_m}}\left( 1+z \right)^{-3/2} .
\label{eq:ztsimp}
\end{equation}
Denoting a derivative with respect to $z$ by a prime, equation
(\ref{eq:diffeqt}) becomes
\begin{equation}
\delta_s^{\prime\prime} +
\frac{1}{2\left( 1+z \right)} \delta_s^{\prime} 
- \frac{3}{\Omega_m H_0^2 \left( 1+z \right)^5} \left( \frac{\partial
\bar{H}}{\partial t} + \bar{H}^2 \right) = - \frac{1}{\Omega_m H_0^2
\left( 1+z \right)^5} \nabla \cdot \left[ \frac{1}{\rho_s} 
\sum_{forces} {\bf F}_s \right] . 
\label{eq:diffeqf}
\end{equation}

We now obtain expressions for the forces.  The force due to gravity, 
when expressed in terms of the potential, $\phi$, is
${\bf F}_s = -\rho_s \nabla\phi $ .
We insert this into equation (\ref{eq:diffeqf}) and use Poisson's
equation which gives 
\begin{equation}
\delta_s^{\prime\prime} + \frac{1}{2\left( 1+z \right)} \delta_s^{\prime} =
- \frac{1}{\Omega_m H_0^2 \left( 1+z \right)^5} \nabla \cdot \left[
\frac{1}{\rho_s} \sum_{forces}\nolimits' {\bf F}_s \right] +
\frac{3}{2\Omega_m} \left( 1+z \right)^{-2} \sum_{species} \Omega_i\delta_i
,
\label{eq:gravity}
\end{equation}
where the primed sum means a sum over all forces except gravity.

Although all types of particles are affected by gravity, we assume that
only the baryons ever obtain a non-zero temperature.  Each
species of baryons experiences a different pressure force which is 
determined by the negative of its pressure gradient,
\begin{equation}
{\bf F}_s=-\nabla P_s 
= -\frac{k_B}{m_s}\left( \rho_s\nabla T_s + T_s\nabla\rho_s  \right) ,
\label{eq:prforce}
\end{equation}
where we have used $m_s$ for the atomic mass and $T_s$ for the temperature
of species $s$.  Dividing equation (\ref{eq:prforce}) by the mass density
and taking the divergence, we see that
\begin{equation}
\nabla\cdot \frac{{\bf F_s}}{\rho_s} = - \frac{k_B}{m_s}\left(
\nabla^2 T_s +T_s\nabla^2\delta_s  \right) 
\label{eq:prlin}
\end{equation}
to first order in the perturbations.  We define $\bar{T}_s$ to be the
temperature in the absence of perturbations, and express the effects of
perturbations in terms of the small quantity $\tau_s$: $T_s =
\bar{T}_s\left( 1+\tau_s\right) $.  Inserting this into equation
(\ref{eq:prlin}) and linearizing once more we see that
\begin{equation}
\nabla\cdot\frac{{\bf F_s}}{\rho_s} = 
- \frac{k_B\bar{T}_s}{m_s}\left( \nabla^2 \tau_s + \nabla^2\delta_s
\right) .
\label{eq:prtaucontrib}
\end{equation}

The temperature of the baryons is determined by
two effects, the coupling of free electrons to the background 
radiation and the adiabatic expansion (Peebles 1993). This gives
\begin{equation}
\frac{dT_s}{dt}=\frac{x}{1+x}\frac{8\sigma_T a T_{\gamma}^4}
{3m_e c}\left( T_{\gamma}- T_s \right) +\frac{2T_s}{3\rho_s}
\frac{d\rho_s}{dt} ,
\label{eq:DTdt}
\end{equation}
where $x$ is the free electron fraction, $\sigma_T$ is the Thomson cross
section, $a$ is the Stefan-Boltzmann constant, $m_e$ is the electron mass,
$c$ is the speed of light, and $T_{\gamma}$ is the temperature of the
cosmic microwave background (CMB).  Switching the independent variable to
$z$ and linearizing, equation (\ref{eq:DTdt}) becomes
\begin{equation}
\frac{dT_s}{dz}
= -\frac{x}{1+x}\frac{8\sigma_T a T_{0}^5\left( 1+z \right)^{5/2}}
{3m_e c H_0 \sqrt{\Omega_m}}
\left( 1 - \frac{T_s}{T_0 \left( 1+z \right)} \right) 
+\frac{2T_s}{\left( 1+z \right)} +\frac{2}{3}T_s\delta_s^{\prime} ,
\label{eq:DTDZ}
\end{equation}
where $T_0=2.73$ K is the CMB temperature at present.

\begin{figure}
\begin{center}
\plotone{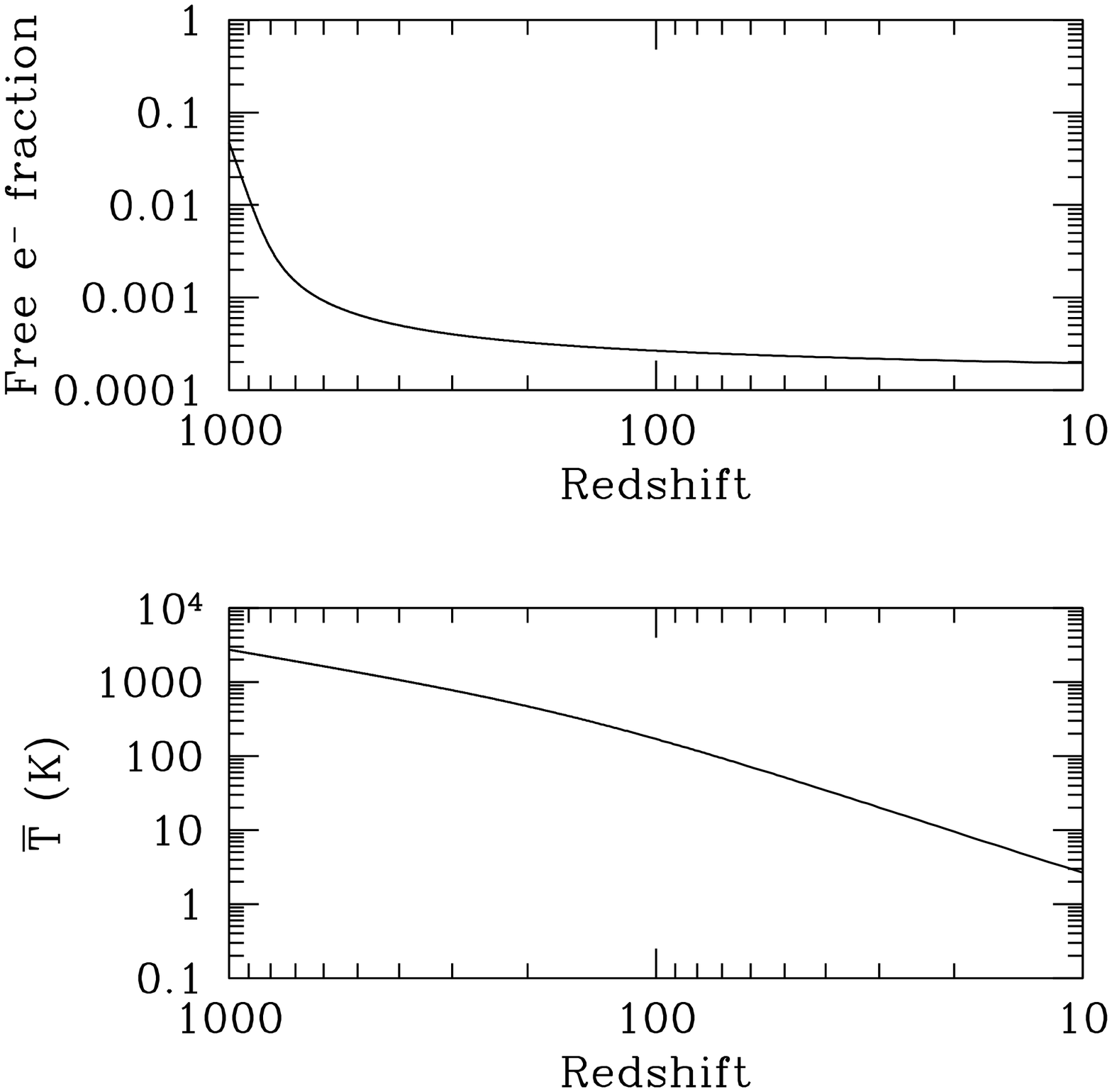} \figcaption[fig1]{Top panel: The free electron fraction
as a function of redshift.  The Universe rapidly approaches neutrality
after $z\approx 1000$.  Bottom panel: The average baryon temperature as a
function of redshift.  For $z>200$, the free electron fraction is still
large enough so that the baryons are coupled to the background radiation,
and so the baryonic temperature decreases like $\left( 1+z \right)$.
For smaller redshifts, the baryon temperature is mainly determined by the
adiabatic expansion, and drops off like $\left( 1+z \right)^{2}$.
\label{fig:xTz}}
\end{center}
\end{figure}

The mean baryon temperature, $\bar{T}_s$, is defined to be a solution of
\begin{equation}
\frac{d\bar{T_s}}{dz}=
-\frac{x}{1+x}\frac{8\sigma_T a T_{0}^5\left( 1+z \right)^{5/2}}
{3m_e c H_0 \sqrt{\Omega_m}}
\left( 1 - \frac{\bar{T}_s}{T_0 \left( 1+z \right)} \right) 
+\frac{2\bar{T}_s}{\left( 1+z \right)} .
\label{eq:Tempzero}
\end{equation}  
This indicates that $\bar{T}_s$ is independent of baryonic species and so
all baryons would be at the same temperature in the absence of
perturbations.  We use the computer program RECFAST (Seager, Sasselov, \&
Scott 1999) to determine the free-electron fraction as a function of
redshift. Then we use equation (\ref{eq:Tempzero}) to compute the mean
baryon temperature.  Both quantities are shown in Figure~\ref{fig:xTz}.

After linearizing  equation (\ref{eq:DTDZ}), we obtain the equations 
for the temperature perturbations, 
\begin{equation}
w^{\prime}_s - 
\frac{x}{1+x}\frac{8\sigma_T a T_{0}^5\left( 1+z \right)^{5/2}}
{3m_e c H_0 \sqrt{\Omega_m}}
\frac{w_s}{\bar{T}_s} =
\frac{x}{1+x}\frac{16\sigma_T a T_{0}^5\left( 1+z \right)^{5/2}}
{9m_e c H_0 \sqrt{\Omega_m}}
\frac{\delta_s}{\bar{T}_s} ,
\label{eq:dwdz}
\end{equation}
where $w_s=\tau_s-\frac{2}{3}\delta_s$.  
With this definition of $w_s$, equation (\ref{eq:prtaucontrib}) becomes
\begin{equation}
\nabla\cdot\frac{{\bf F_s}}{\rho_s} = 
-\frac{k_B \bar{T}}{m_s}\left( \frac{5}{3}\nabla^2 \delta_s +
\nabla^2 w_s \right) . 
\label{eq:prfinal}
\end{equation}
Because of the spatial derivatives, we are forced to work in comoving
Fourier space.  However, for simplicity of notation we will continue to
simply write $\delta_s$ and $w_s$, although from now on we use these
symbols to represent the Fourier components of the perturbations.

Finally we take into account the frictional term arising from inter-species
collisions.  Physically, this force is non-zero only if two species have
different bulk velocities; thus, the force on one species due to a second
species may be written as ${\bf F}_1= \kappa\left( {\bf x},t\right) \left(
{\bf v}_2 - {\bf v}_1 \right) $ (Burgers 1969).  Since the difference in
bulk velocities is already linear in the perturbations, we only need to
calculate the coefficient $\kappa$ to zeroth order for the linearized
equations of motion.  This allows us to calculate $\kappa$ under the
assumption that all baryons share the same temperature.  In general, the
frictional force per unit volume acting on species $i$ due to species $j$
is obtained by multiplying the number of collisions per unit time per unit
volume by the momentum transfer per collision.  We multiply this by the
probability that species $i$ and $j$ have initial momentum ${\bf p}_i$ and
${\bf p}_j$ and that the center of mass scattering angle is $\chi$.
Finally, we integrate over all parameters,
\begin{equation}
{\bf F}_{ij}= \int d^3{\bf p}_i \int d^3{\bf p}_j \int d\Omega\  
n_i n_j \sigma_{ij} \left| {\bf v}_i - {\bf v}_j \right| 
\Delta {\bf p}_i \left( {\bf p}_i , {\bf p}_j ,
\chi \right) f_i\left( {\bf p}_i \right) 
f_j\left( {\bf p}_j \right) \frac{1}{\sigma_{ij}}
\frac{d \sigma_{ij}}{d\Omega}\left( {\bf p}_i , 
{\bf p}_j , \chi \right) .
\label{eq:collint}
\end{equation}

\begin{figure}
\begin{center}
\plotone{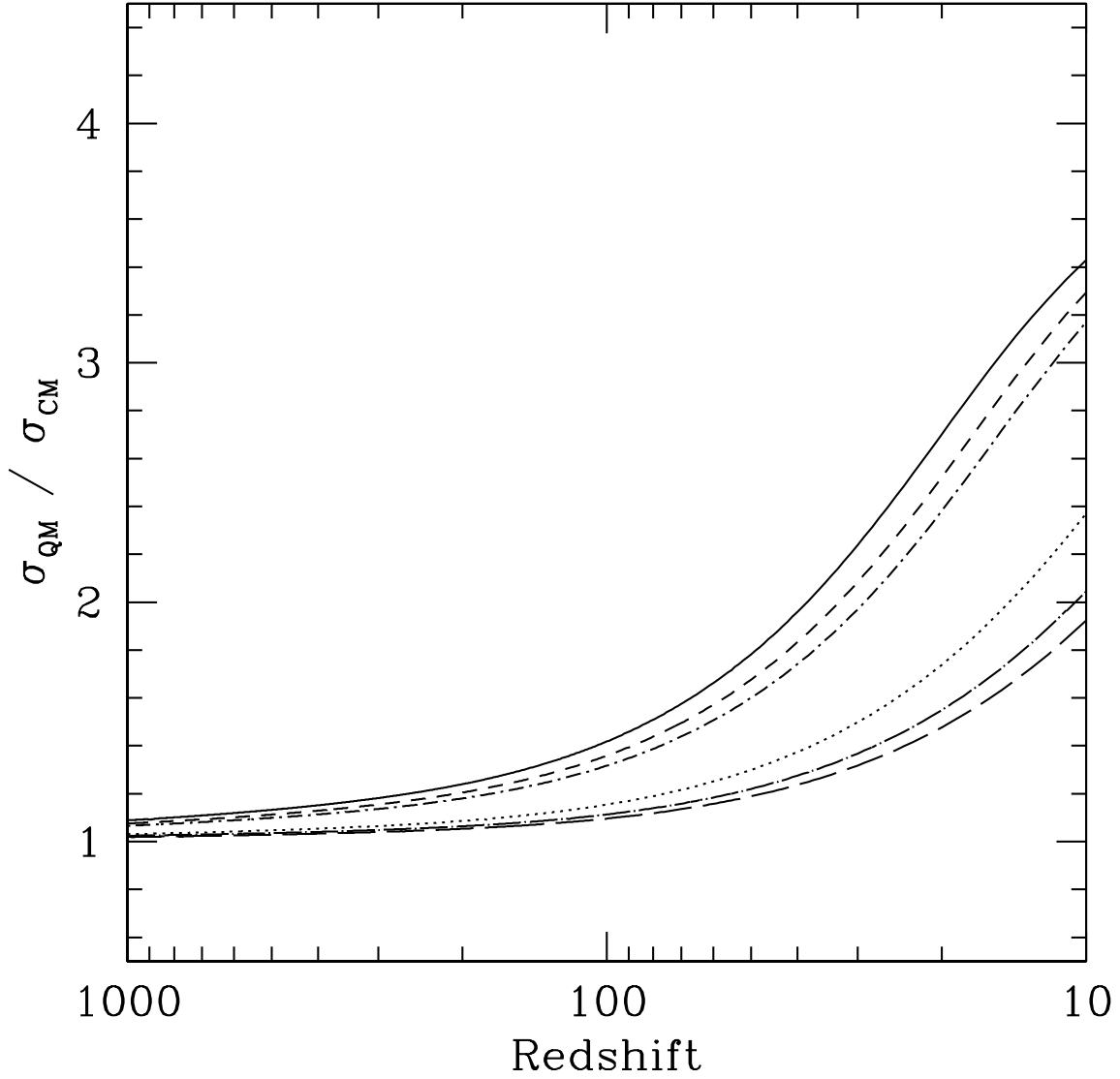} \figcaption[fig1]{The inter-particle cross section, in
units of the classical hard-sphere cross section.  Solid line is
$^1$H-$^4$He, dotted is $^1$H-$^7$Li, short dash is $^1$H-D, long dash is
$^4$He-$^7$Li, dot-short dash is $^4$He-D, and dot-long dash is $^7$Li-D.
This quantum-mechanical correction factor depends on the atomic radii, the
reduced mass, and the temperature.
\label{fig:xsect}}
\end{center}
\end{figure}

We assume that the collisions are elastic and that the atoms can be
approximated by hard spheres.  We take each species to have a Maxwellian
velocity distribution.  Following the integration equation
(\ref{eq:collint}) becomes
\begin{equation}
{\bf F}_{ij}= -\frac{16 m_{ij}}{3\sqrt{2\pi}}\sqrt{\frac{k_B \bar{T}}{m_{ij}}}
n_i n_j \sigma_{cl}\left( {\bf v}_i - {\bf v}_j \right)
\int^{\infty}_0 dx x^5 e^{-x^2} Q_d\left( x \sqrt{\frac{k_B \bar{T}}{m_{ij}}}
\right) .
\label{eq:collforce}
\end{equation}
Here, $m_{ij}=m_i m_j/(m_i+m_j)$ is the reduced mass, $\sigma_{cl}$ is the
classical hard-sphere cross section, $n_i=\rho_i/m_i$, and the integral is
a quantum mechanical correction (Massey, Burhop, \& Gilbody 1969).  The
function $Q_d\left( q \right)$ is given by
\begin{equation}
Q_d = \frac{1}{q^2}\sum_{l=0}^{\infty}\left[ 
4\left( 2l+1 \right) \sin^2\eta_l - 2 \left( l+1 \right) \left(
1-\cos 2\eta_l - \cos 2\eta_{l+1} + \cos 2\left( \eta_l - \eta_{l+1} \right)
\right) \right] .
\label{eq:qmcor}
\end{equation}
The integration variable $x$ is related to $q$ by \( q=x a \sqrt{{2 m_{ij}
k_B \bar{T}}/{\hbar^2}} \) , and the phase shifts are given by \( \eta_0 =
-q \) and \( \tan\eta_i = \left( -1 \right)^{i-1} { J_{i+1/2}\left( q
\right)}/ {J_{-i-1/2}\left( q\right)} \) , where the functions $J_n\left(
q\right)$ are Bessel functions of order $n$.  The distance scale $a$
represents the sum of the radii of two colliding atoms; we adopt the values
$r_H=r_{D}=0.53$\AA~, $r_{He}=0.31$\AA~, and $r_{Li}=1.45$\AA~ (Clementi,
Raimondi, \& Reinhardt 1963; Slater 1964).  We evaluate the integral in
equation (\ref{eq:collforce}) numerically and plot the results in
Figure~\ref{fig:xsect} for each type of inter-species collision.  Quantum
mechanics becomes more important as the temperature decreases but the
correction is never greater than a factor of four.  For brevity, we
henceforth denote this integral by $I_{ij}\left( z \right)$.

We now summarize the equations governing each species. For dark matter
equation (\ref{eq:diffeqf}) becomes
\begin{equation}
\delta_{dm}^{\prime\prime} + \frac{1}{2}\left( 1+z \right)^{-1}
\delta_{dm}^{\prime}=\frac{3}{2} \left( 1+z \right)^{-2}
\sum_{species}\frac{\Omega_i}{\Omega_m} \delta_i .
\label{eq:dmdeq}
\end{equation}
For each baryonic species there are also pressure and collisional friction
terms, and so
\begin{eqnarray}
\delta_{i}^{\prime\prime} + \frac{1}{2}\left( 1+z \right)^{-1}
\delta_{i}^{\prime} & = & \frac{3}{2} \left( 1+z \right)^{-2}
\sum_{j=all\ species}\frac{\Omega_j}{\Omega_m} \delta_j \label{eq:bdeq} \\  
                    &   &  -\frac{k_B \bar{T}}{m_i\Omega_m H_0^2}
\left( 1+z \right)^{-3} k^2 \left( \frac{5}{3}\delta_i +w_i\right) \nonumber \\
                    &   &   + \frac{H_0}{G}
\sqrt{\frac{2 k_B \bar{T}}{\Omega_m \pi^3}}\sum_{j\neq i}
\frac{\Omega_j\sigma^{ij}_{cl}}{\sqrt{m_i m_j 
\left( m_i + m_j \right)}}
\left( 1+z \right)^{1/2} I_{ij}\left( z \right)
\left( \delta^{\prime}_i - \delta^{\prime}_j \right) \nonumber ,
\end{eqnarray}
where $k$ is the comoving wave number.  From equation (\ref{eq:bdeq}) we
can easily find the differential equations satisfied by the differences
relative to hydrogen, $\Delta_i=\delta_i - \delta_H$:
\begin{eqnarray}
\Delta_{i}^{\prime\prime} + \frac{1}{2}\left( 1+z \right)^{-1}
\Delta_{i}^{\prime} & = & \frac{k_B \bar{T}}{\Omega_m H_0^2}
\left( 1+z \right)^{-3} k^2 \left[ \frac{5}{3}\left( \frac{\delta_H}{m_H}
+\frac{\Delta_i-\delta_H}{m_i}\right) 
+ \left( \frac{w_H}{m_H} - \frac{w_i}{m_i}
\right) \right]  \\
                    &   &   - \frac{H_0}{G}
\sqrt{\frac{2 k_B \bar{T}}{\Omega_m \pi^3}}\sum_{j\neq H}
\frac{\Omega_j\sigma^{H-j}_{cl}}{\sqrt{m_H m_j 
\left( m_H + m_j \right)}}
\left( 1+z \right)^{1/2} I_{H-j}\left( z \right)
\Delta^{\prime}_j  \nonumber  \\
                    &   &   + \frac{H_0}{G}
\sqrt{\frac{2 k_B \bar{T}}{\Omega_m \pi^3}}\sum_{j\neq i}
\frac{\Omega_j\sigma^{ij}_{cl}}{\sqrt{m_i m_j 
\left( m_i + m_j \right)}}
\left( 1+z \right)^{1/2} I_{ij}\left( z \right)
\left( \Delta^{\prime}_j - \Delta^{\prime}_i \right) \nonumber .
\label{eq:diffdeq}
\end{eqnarray}

\section{Initial Conditions}
\label{sec:inicond}

We consider dark matter perturbations initially of the spherical
top-hat form
\( \delta_{dm}\left( {\bf x}, z_{ini} \right) = 
   \delta_{ini} \theta\left( R - \left| {\bf x} \right| \right) \), 
which has the Fourier components 
\begin{equation}
\delta_{dm}\left( {\bf k},z_{ini}\right) = 
\frac{4\pi\delta_{ini}}{k^3}\left( \sin k R - k R\cos k R \right) .
\label{eq:tophatf}
\end{equation}
Here, $\delta_{ini}$ and $R$ are constants and $\theta$ represents the step
function.  We take the initial values of $\Delta_{He}$, $\Delta_{Li}$,
$\Delta_{D}$ and their derivatives to be zero, but we would like to choose
the initial values of $\delta_{dm}^{\prime}$, $\delta_H$, and
$\delta_H^{\prime}$ that correspond to the most rapidly growing mode. To
gain some insight into how to make this choice, we consider a simplified
problem where baryonic diffusion is negligible.  We approximate the
baryonic temperature by the temperature of the background radiation (valid
for $z\gtrsim 200$; see Figure~\ref{fig:xTz}).  Equation (\ref{eq:bdeq})
for the baryon overdensity becomes
\begin{eqnarray}
\delta_{b}^{\prime\prime} + \frac{1}{2}\left( 1+z \right)^{-1}
\delta_{b}^{\prime} & = & \frac{3}{2} \left( 1+z \right)^{-2}
\left( \frac{\Omega_{dm}}{\Omega_m} \delta_{dm} 
+ \frac{\Omega_b}{\Omega_m}\delta_b \right) \label{eq:bnodiff} \\  
                    &   &  -\frac{k_B T_0}{\mu\Omega_m H_0^2}
\left( 1+z \right)^{-2} k^2 \delta_b . \nonumber 
\end{eqnarray}
We use this equation to express $\delta_{dm}$, differentiate, and insert
the results into equation (\ref{eq:dmdeq}).  The outcome is a fourth order
differential equation for $\delta_b$.  It has four independent, power law
solutions which we write as \( \delta_b = \sum_{i=1}^4 A_i \left( 1+z
\right) ^{p_i} \).  The four $A_i$ are the four arbitrary constants
associated with a linear fourth order equation.  The four $p_i$ are given
by
\begin{equation}
p_i = \left\{  \begin{array}{l}
 \frac{1}{4}+\sqrt{\frac{13}{16}-\frac{\hat{k}^2}{2}+\frac{1}{2}
\sqrt{\left( \hat{k}^2 - \frac{3}{2}\right)^2+6\hat{k}^2
\frac{\Omega_{dm}}{\Omega_m}}}  \\
 \frac{1}{4}+\sqrt{\frac{13}{16}-\frac{\hat{k}^2}{2}-\frac{1}{2}
\sqrt{\left( \hat{k}^2 - \frac{3}{2}\right)^2+6\hat{k}^2
\frac{\Omega_{dm}}{\Omega_m}}}  \\
 \frac{1}{4}-\sqrt{\frac{13}{16}-\frac{\hat{k}^2}{2}+\frac{1}{2}
\sqrt{\left( \hat{k}^2 - \frac{3}{2}\right)^2+6\hat{k}^2
\frac{\Omega_{dm}}{\Omega_m}}}  \\
 \frac{1}{4}-\sqrt{\frac{13}{16}-\frac{\hat{k}^2}{2}-\frac{1}{2}
\sqrt{\left( \hat{k}^2 - \frac{3}{2}\right)^2+6\hat{k}^2
\frac{\Omega_{dm}}{\Omega_m}}} ,  
\end{array} 
\right. 
\label{eq:inipower}
\end{equation}
where $\hat{k}^2$ is the squared comoving wavenumber normalized by
$({\mu\Omega_m H_0^2}/{k_B T_0})$.  The resulting expression for the dark
matter overdensity is
\begin{equation}
\delta_{dm} = \sum_{i=1}^4 c_i A_i \left( 1+z \right)^{p_i} ,
\label{eq:dminisol}
\end{equation}
with the constants $c_i$ given by
\begin{equation}
c_i = p_i^2\left( \frac{2\Omega_m}{3\Omega_{dm}} \right)
- p_i \left( \frac{\Omega_m}{3\Omega_{dm}} \right)
+\frac{2\hat{k}^2\Omega_m -3 \Omega_b}{3\Omega_{dm}} .
\label{eq:cofk}
\end{equation}

Motivated by this solution, we choose our initial conditions so that the
perturbation will grow like the most negative $p_i$ (i.e., the third one
listed in equation~\ref{eq:inipower}).  We let the three $A_i$
corresponding to all but the largest growth rate be zero, and let the final
$A_i$ be determined by the tophat distribution (equation \ref{eq:tophatf}).
The remaining initial conditions may now be written in terms of the known
$c_i$ and $p_i$.

\section{Calculations}
\label{sec:calc}
\subsection{Evolution of the k-Modes}
\label{sec:kmodes}

We define a new set of variables, 
\begin{equation}
\hat{\delta}_s=\frac{k^3}
{4\pi \delta_{ini} \left( \sin kR - kR\cos kR \right) }
\delta_s ,
\label{eq:normdeldef}
\end{equation}
\begin{equation}
\hat{\Delta}_s=\frac{k^3}
{4\pi \delta_{ini} \left( \sin kR - kR\cos kR \right) }
\left( \delta_s - \delta_H \right),
\label{eq:ncapddef}
\end{equation}
\begin{equation}
\hat{w}_s=\frac{k^3}
{4\pi \delta_{ini} \left( \sin kR - kR\cos kR \right) } w_s . 
\label{eq:normwdef}
\end{equation}
These quantities are special because they not only satisfy 
the same differential equations as the $\delta$ and $w$,
but also satisfy simpler initial conditions,
\( \hat{\delta}_{dm} = 1 \) ,
\( \hat{\delta}_{dm}^{\prime} = \frac{p_3 } {1+z_{ini}} \) ,
\( \hat{\delta}_b = c_3^{-1} \) , and 
\( \hat{\delta}_b^{\prime} = \frac{p_3 } 
{c_3 \left( 1 + z_{ini} \right) } \) . 
The parameters $\left( \delta_{ini} , R \right) $ have disappeared from the
problem.  Therefore we only evolve each normalized $k$-mode once; we obtain
the physical overdensity for a given $\delta_{ini}$ and $R$ by multiplying
each $k$-mode by the appropriate factor from equation
(\ref{eq:normdeldef}).

We have set $z_{ini}=10^3$ following cosmological recombination.  The final
redshift is marked by the first of two possible events. Our equations are
applicable in the linear regime, so we must stop if the object becomes
nonlinear.  At a transition to nonlinearity, the density would be greatly
enhanced which would result in the suppression of diffusion.  Therefore, we
expect diffusion to be effective in the linear regime where our equations
are valid.  The second stopping condition relates to the reionization of
the Universe.  At that time, large Coulomb cross sections between charged
particles are able to effectively halt diffusion.  In conclusion, we
consider the case where the object becomes nonlinear after reionization and
adopt $z_{reion}=8$ (Haiman \& Loeb 1998, 1999a,b; Gnedin \& Ostriker 1997;
Chiu \& Ostriker 2000; Gnedin 2000) as the stopping point of our diffusion
calculation.  Since the collapsed fraction of the Universe is small at
$z=8$ (Barkana \& Loeb 2001), most of the material within galaxies that
form much later (at $z\la 3$) was still in the linear regime at $z\ga 8$,
where our diffusion calculation is adequate.

\begin{figure}
\begin{center}
\plotone{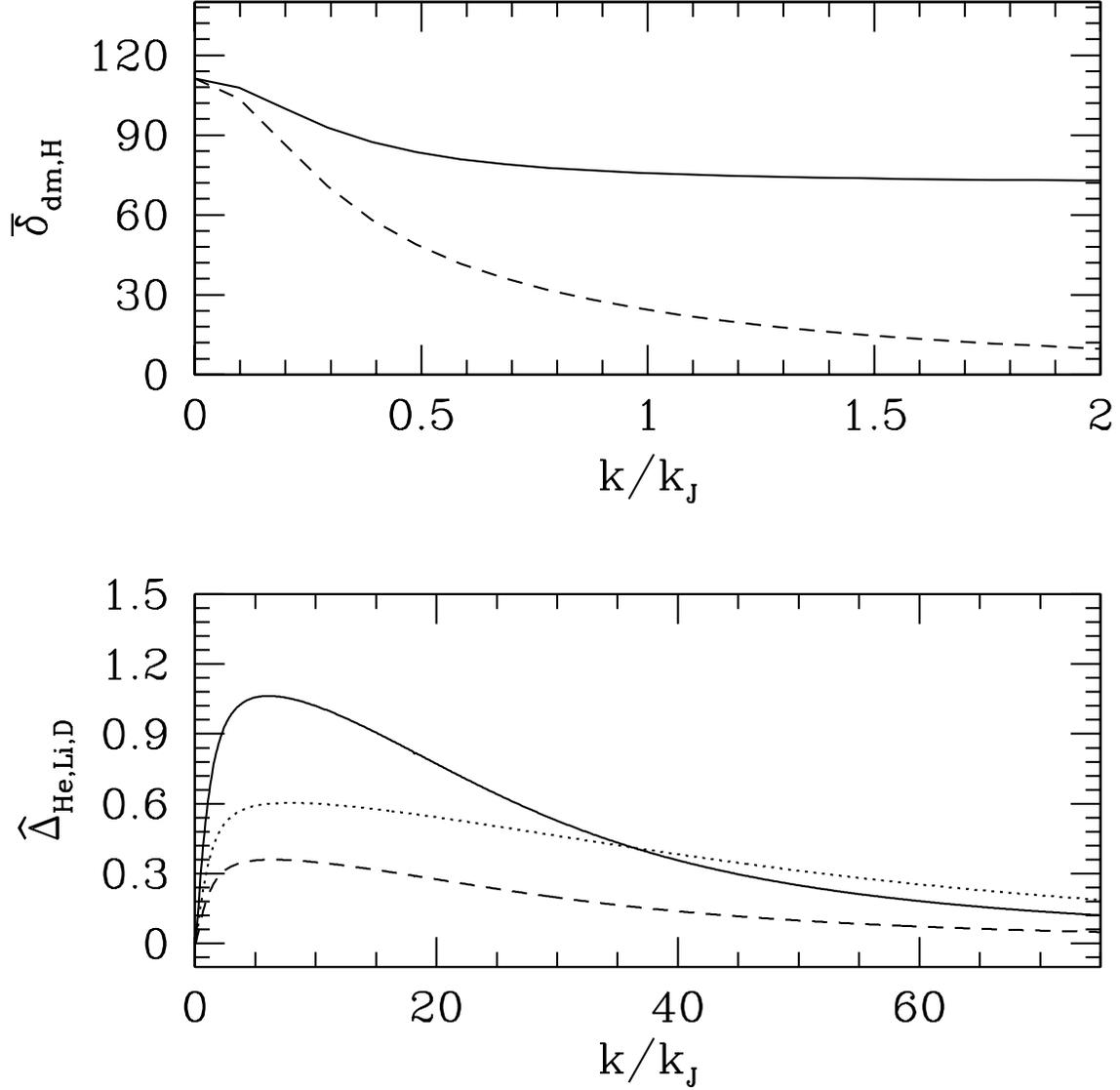} \figcaption[fig1]{ {\it Top panel:} The normalized
dark matter (solid) and hydrogen (dashed) overdensities.  For small $k$,
both are amplified.  But as $k$ increases, pressure forces inhibit baryonic
modes from growing. {\it Bottom panel:} The normalized differences relative
to Hydrogen, $\hat{\Delta}_x$.  The solid line has $x=^4$He, the dotted
line has $x=^7$Li, and the dashed line has $x=$D.  The relatively heavy
mass and small atomic radius of $^4$He make it particularly susceptible to
diffusion.
\label{fig:nplots}}
\end{center}
\end{figure}

Several results of having carried out this prescription are shown in
Figure~\ref{fig:nplots}.  All length scales have been measured in terms of
$k_J^{-1}$, defined in equation (\ref{eq:k_J}).  This expression involves
the baryonic temperature which is set by the adiabatic expansion for
redshifts less than $z_t\approx 138$, \( T\left( z\right)=T_0\left(
1+z\right)^2\left( 1+z_t\right)^{-1} \); as discussed in
\S~\ref{sec:baseq}.  At $z_{reion}$, the Jeans wavenumber therefore obtains
the value
\begin{equation}
k_J= 1040\ {\rm Mpc}^{-1}\left[ \left(\frac{9}{1+z}\right) 
\left(\frac{\Omega_m h^2}{0.147}\right) 
\left(\frac{2.73\ {\rm K}}{T_0}\right) \right]^{1/2},
\label{eq:kJnumber}
\end{equation}
which is the value used in Figure~\ref{fig:nplots}.  We see that the dark
matter overdensity has evolved approximately proportionally to $\left( 1+z
\right)^{-1}$, as might have been expected since gravity is the only force
applicable to dark matter.  On large scales corresponding to $k\lesssim
k_J$, the baryons follow the dark matter; however, on small scales
corresponding to $k\gtrsim k_J$, pressure forces prevent baryonic modes
from collapsing.  This leads to the slight decline in dark matter
amplification near the Jeans scale and the dramatic decline in the
amplification of baryonic modes there.  In the lower panel of Figure 3, we
see that the perturbations of the different baryonic species do not
experience exactly the same amplification.  At $k\ll k_J$, pressure is
negligible for all species, and the modes of each are amplified. Since
hydrogen is the lightest species, it reaches its Jeans scale at the lowest
$k$.  Thus, all of the $\hat{\Delta}_x$ become positive at approximately
the same wavenumber.  As $k$ increases, we begin to reach the Jeans scales
of the heavier species.  Then the growth of $k$-modes of this second
species is also suppressed, and the $\hat{\Delta}_x$ approach zero as $k$
gets large.  However, Figure~\ref{fig:nplots} indicates that the overall
magnitude of diffusion is not solely determined by the Jeans scale of a
species, or else the effect would have been largest for $^7$Li.  This is
because of the frictional term which serves to suppress diffusion, which is
proportional to the square of the radius of an atom and therefore the
greatest suppression is associated with the largest atoms.

\begin{figure}
\begin{center}
\plotone{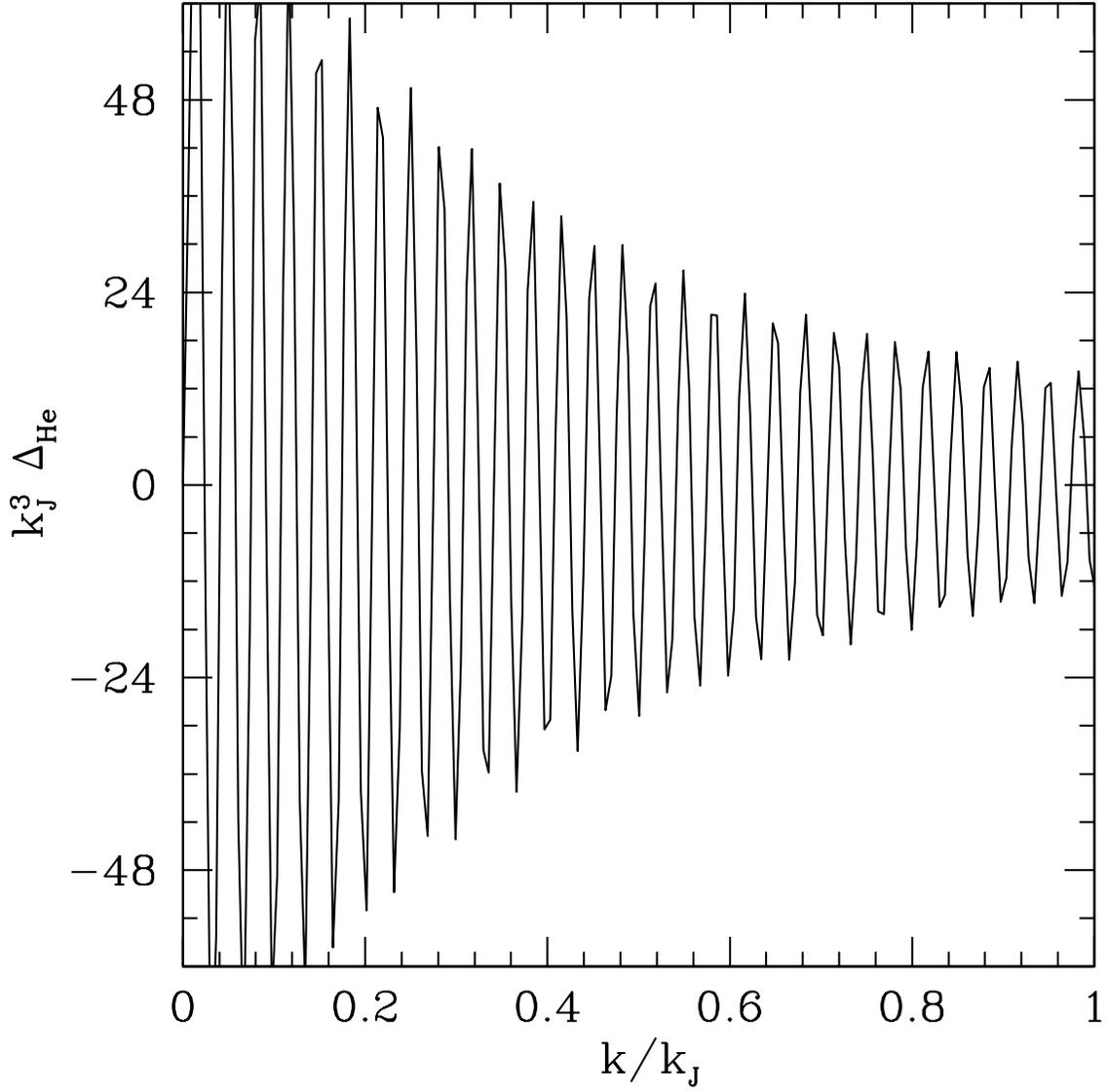} \figcaption[fig1]{The difference $\Delta_{He}$, for
an overdense region with a mass of $10^9{\rm M}_{\odot}$.  This plot is
obtained by interpolating the smoothly varying normalized overdensities,
and then multiplying each mode by its rapidly varying normalization factor.
\label{fig:delMnine}}
\end{center}
\end{figure}

The normalized overdensities in Figure~\ref{fig:nplots} were obtained by
evolving 1025 $k$-modes, equally spaced between with $0\leq k/k_J \leq
100$.  To obtain the physical overdensities, each mode must be multiplied
by its normalization factor.  The normalization factor depends on the
object assembly radius $R$, and is fairly rapidly varying compared to the
normalized overdensities.  To get an accurate determination of the physical
overdensity, we interpolate the normalized overdensity so that we sample it
at a rate larger than the frequency of the normalization factor.  We then
use equation (\ref{eq:normdeldef}) to obtain the physical $k$-modes.  One
such sampling is shown in Figure~\ref{fig:delMnine}.  The value of $R$ used
for this plot corresponds to an enclosed mass of $M=10^9
M_{\odot}=\rho_{crit} \Omega_m\frac{4}{3}\pi R^3$.  The relative errors
incurred in interpolation are typically $\sim 10^{-6}$, and seldom exceed
$10^{-5}$ with the 1025 initial points.  The normalization factor is also
proportional to the initial dark matter perturbation, $\delta_{ini}$, which
we express in terms of the collapse redshift of the object ($z_{coll}$)
through the approximate relation \mbox{\( \delta_{ini} \left( 1+z_{ini}
\right) = \delta_{coll}\left( 1+z_{coll} \right) \) .}  We take
$\delta_{coll}=1.69$ for the overdensity of collapse and, unless otherwise
stated, $z_{coll}=2.5$ for the redshift of collapse.  Such objects would be
just on the fringe of nonlinearity at $z_{reion}=8$.  Since diffusion
within all objects was shut off at $z_{reion}$, the effects of diffusion
would be decreased for objects that collapsed later.

\subsection{Predicted Statistics}
\subsubsection{Mean Abundances}
Next we quantify the extent to which element abundances in collapsed
objects differ from their primordial values.  Abundances of $^4$He are 
commonly reported in terms of the mass fraction,
\begin{equation}
Y = \frac{\bar{\rho}_{He}\left( 1 + \delta_{He} \right) }
{\bar{\rho}_{He}\left( 1 + \delta_{He} \right) +  
\bar{\rho}_H\left( 1 + \delta_{H} \right) } ,
\label{eq:ymdef}
\end{equation}
where we have neglected the contributions of all species but $^4$He and
$^1$H to the mass of the Universe. After linearizing this equation,
we see that
\begin{equation}
\frac{Y-\bar{Y}}{\bar{Y}}= \left( 1 - \bar{Y} \right) 
\left( \delta_{He} - \delta_H \right) .
\label{eq:ymdev}
\end{equation}
In contrast, the abundances of $^7$Li and D are typically reported in terms
of their number densities relative to hydrogen.  In this case,
\begin{equation}
\frac{X_{L,D}-\bar{X}_{L,D}}{\bar{X}_{L,D}}= 
\left( \delta_{Li,D} - \delta_{H} \right) .
\label{eq:lidedev}
\end{equation}

To see how the element abundances in collapsed objects differ, on average,
from their primordial values, we average equations (\ref{eq:ymdev})-
(\ref{eq:lidedev}) over the original sphere of comoving radius $R$,
\begin{equation}
\Delta_R = \frac{3}{4\pi R^3}\int_{sphere} d^3{\bf x}
\int \frac{d^3{\bf k}}{\left( 2\pi \right)^3} 
\Delta_{\bf k} e^{i {\bf k}\cdot {\bf x}} .
\label{eq:mdeldef}
\end{equation}
After integrating, we see that
\begin{equation}
\Delta_R = \frac{3}{2\pi^2 R^3}\int_0^{\infty}
\frac{dk}{k}\Delta_k \left( \sin kR -
kR \cos kR \right) .
\label{eq:mdelint}
\end{equation}
We carry out this final integration numerically.  As described in
\S~\ref{sec:kmodes}, we evolve 1025 $k$-modes between 0 and 100$k_J$ to get
the $\hat{\Delta}_x$ (where $x$ could denote any of $^4$He, $^7$Li, D).  We
interpolate to get values for $\hat{\Delta}_x$ at a fixed number of points,
and then multiply each point by its normalization.  We then evaluate the
integral in equation (\ref{eq:mdelint}) with these points.  Then we repeat
the procedure, but with twice as many interpolated points.  If the two
results differ by less than a fraction of $5\times 10^{-3}$, we stop and
accept the result.

\begin{figure}
\begin{center}
\plotone{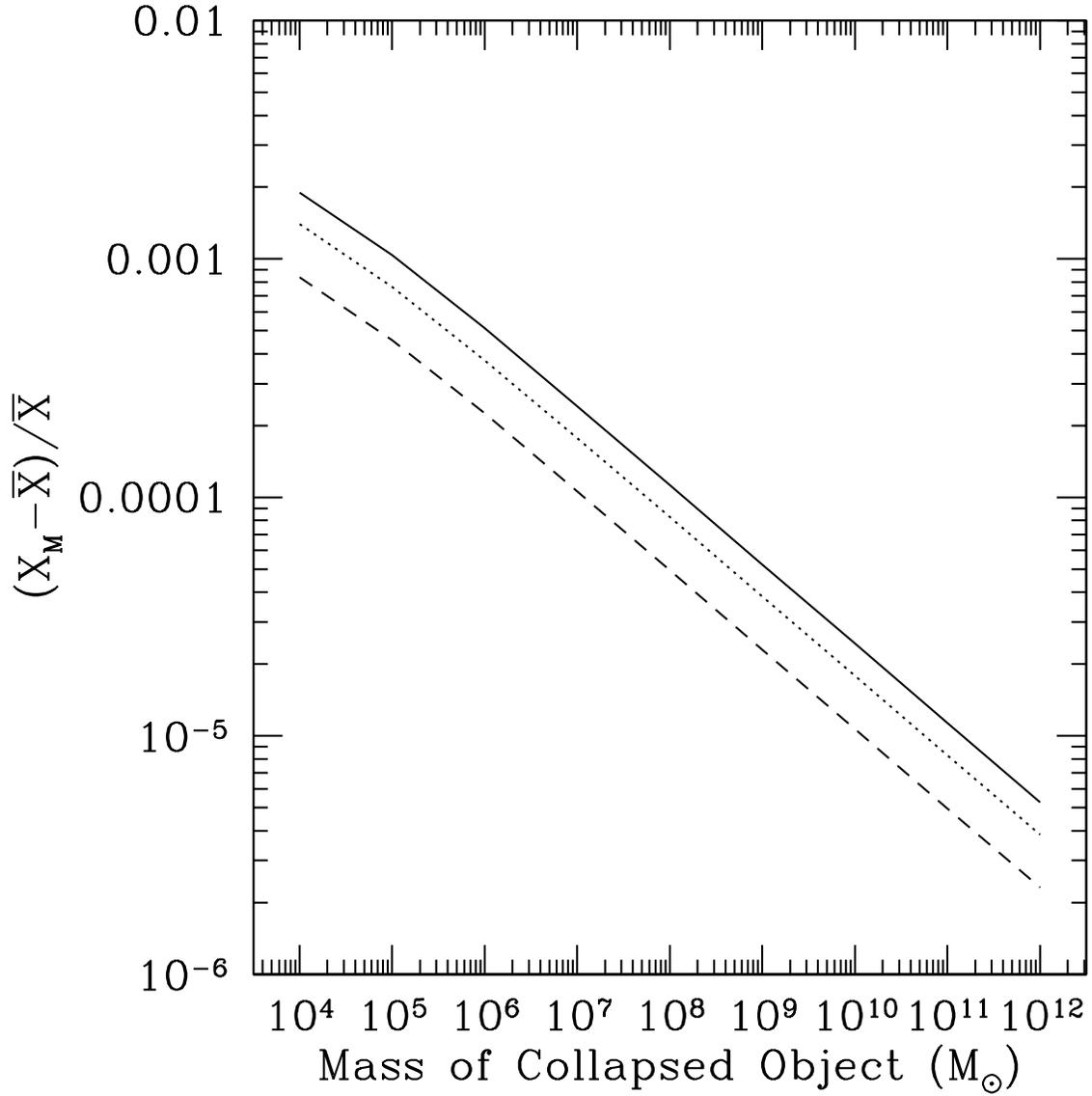} \figcaption[fig1]{ Fractional deviations from the
primordial abundances for $z_{reion}=8$.  The solid line is the fractional
deviation for $^4$He, the dotted line is for $^7$Li, and the dashed line is
for D.
\label{fig:meandev}}
\end{center}
\end{figure}

Figure~\ref{fig:meandev} illustrates the dependence of these deviations on
the size of the collapsed object.  Instead of measuring the size of the
object by its assembly radius, $R$, we instead measure it in terms of its
mass defined by
\begin{equation}
M=\rho_{crit}\Omega_m \frac{4}{3}\pi R^3 .
\label{eq:mrrel}
\end{equation}
We see that the deviation from the primordial mean shrinks as the collapsed
object increases in size approximately according to a power law with slope
$-0.33$.  A simple argument yields the same result: assuming that the
collapsing region starts out with primordial abundances throughout its
volume and that diffusion occurs through the boundary of the region, the
deviations from primordial abundances should fall off like $1/R$, or
$M^{-1/3}$.

\begin{figure}
\begin{center}
\plotone{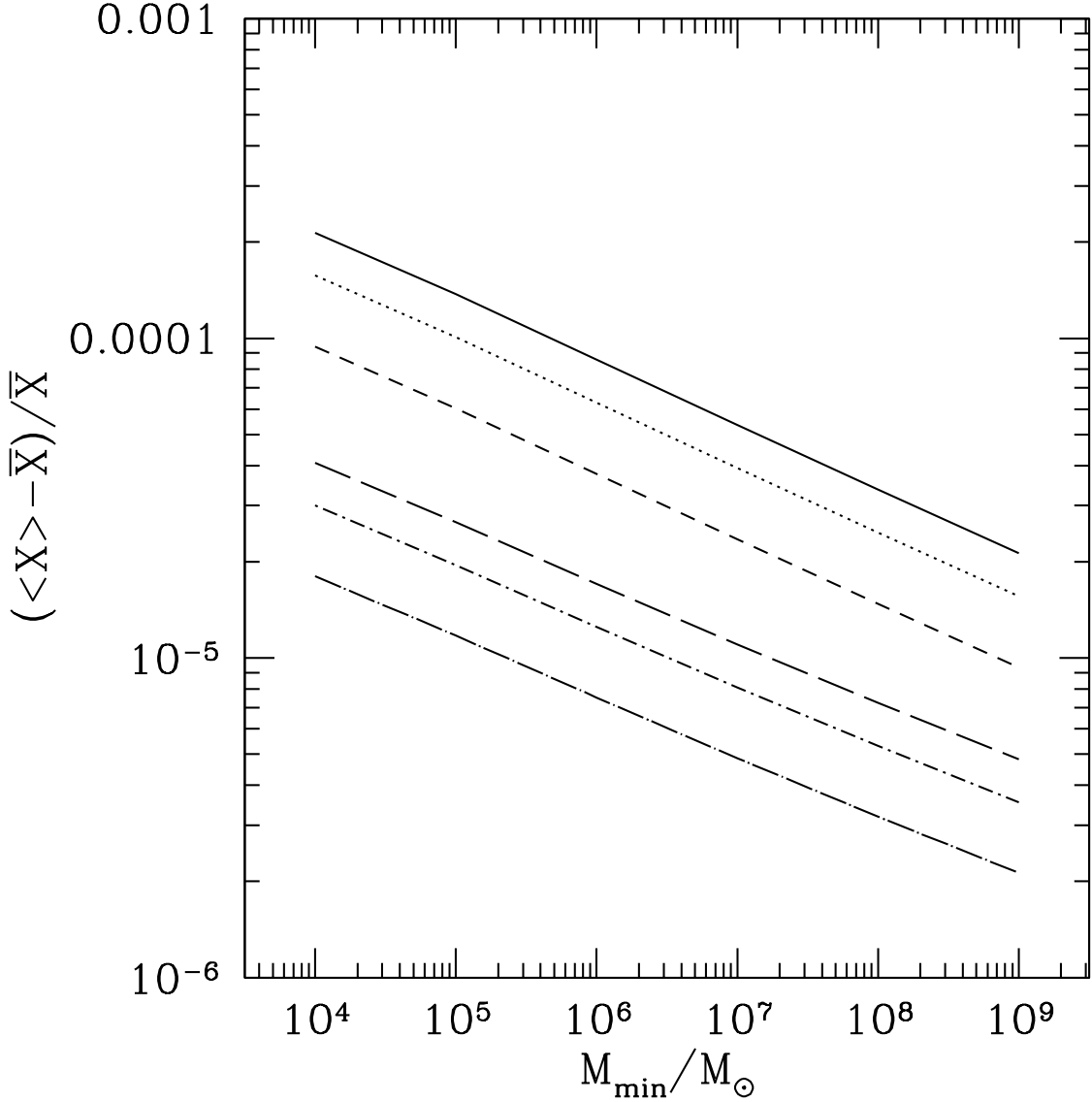} \figcaption[fig1]{The fractional deviations averaged
over all collapsed objects in the Universe, as a function of the minimum
allowed mass scale for galaxies.  We show two cases for the collapse
redshift of the objects. For the first case with $z_{coll}=2.5$, the solid
line corresponds to the fractional deviation for $^4$He, the dotted line to
$^7$Li, and the short-dashed line to D.  For the second case with
$z_{coll}=0$, the long-dashed line corresponds to $^4$He, the dot-short
dashed line to $^7$Li, and the dot-long dashed line to D.
\label{fig:unimean}}
\end{center}
\end{figure}

Now that element abundances for particular collapsed objects are known, we
proceed by averaging these abundances over all objects in the Universe that
have collapsed by a particular redshift.  This statistic is approximated by
using a Press-Schechter mass function; for $^4$He,
\begin{equation}
\left< Y \right> = \frac{\int_{M_{min}}^{\infty} Y_M M \frac{dn}{dM} dM}
{\int_{M_{min}}^{\infty} M \frac{dn}{dM} dM} ,
\label{eq:uniavg}
\end{equation}
where $dn/dM$ is the number density of objects of mass $M$ that have
collapsed at a given redshift and $Y_M$ is the averaged $^4$He mass
fraction within an object of mass $M$.  Equation (\ref{eq:uniavg}) is also
appropriate for the $^7$Li and D abundances if we replace $Y$ with
$X_{L,D}$.  Note also that there is a minimum mass scale of galaxies in
which star formation is possible that we denote by $M_{min}$ (Abel 1995;
Tegmark et~al. 1997).  We use a code developed by Eisenstein \& Hu (1999)
and Barkana (2000) to calculate $dn/dM$ at a redshift of $2.5$.  The
averaged abundances differ from primordial by the amounts shown in
Figure~\ref{fig:unimean}.  We also consider a collapse redshift of
$z_{coll}=0$; in this case Figure~\ref{fig:unimean} shows that the
abundance deviations decrease somewhat.  This is not only due to the
resulting decrease in $\delta_{ini}$, but also due to the dependence of
$dn/dM$ on collapse redshift.  At higher $z$, $dn/dM$ tends to favor lower
mass scales where the abundance deviations are larger
(Figure~\ref{fig:meandev}).

\subsubsection{Fluctuations of Abundances Within Objects}

The element abundances of collapsed objects can differ from primordial
abundances in two ways.  First, the abundances of an object could be
uniformly different from the primordial values. This possibility was
already discussed in the previous section.  Second, if the element
abundances within an object vary greatly from point to point, differences
from the primordial abundances may be observed because of the fluctuations.
We consider this latter effect by computing the standard deviation of the
abundances within a given object.  The lowest-order formula for this
quantity is
\begin{equation}
\sigma_Y^2=\bar{Y}^2\left( 1- \bar{Y} \right) ^2
\sigma_{\Delta_Y}^2 
\label{eq:hesigdef}
\end{equation}
for $^4$He and 
\begin{equation}
\sigma_X^2=\bar{X}^2\sigma_{\Delta_X}^2 
\label{eq:lidesigdef}
\end{equation}
for $^7$Li and D.  We have already calculated the mean values of the
abundances, so now we only need the mean squares.  We proceed as above, but
this time we obtain a double integral that must be evaluated numerically,
\begin{equation}
\left( \Delta^2 \right)_M = \frac{3}{8\pi^4 R^3}
\int_0^{\infty} dk k \Delta_k \int_0^{\infty} dl l \Delta_l
\left( \frac{\sin\left( k-l \right) R}{k-l} -
\frac{\sin\left( k+l\right) R}{k+l} \right) .
\label{eq:msdelp}
\end{equation}
The integration region is the first quadrant of the $k-l$ plane.
Since the integrand is symmetric with respect to the interchange 
of $k$ and $l$, it is equal to the integral
\begin{equation}
\left( \Delta^2 \right)_M = \frac{3}{4\pi^4 R^3}
\int_0^{\infty} dk k \Delta_k \int_0^{k} dl l \Delta_l
\left( \frac{\sin\left( k-l \right) R}{k-l} -
\frac{\sin\left( k+l\right) R}{k+l} \right) .
\label{eq:msdel}
\end{equation}

\begin{figure}
\begin{center}
\plotone{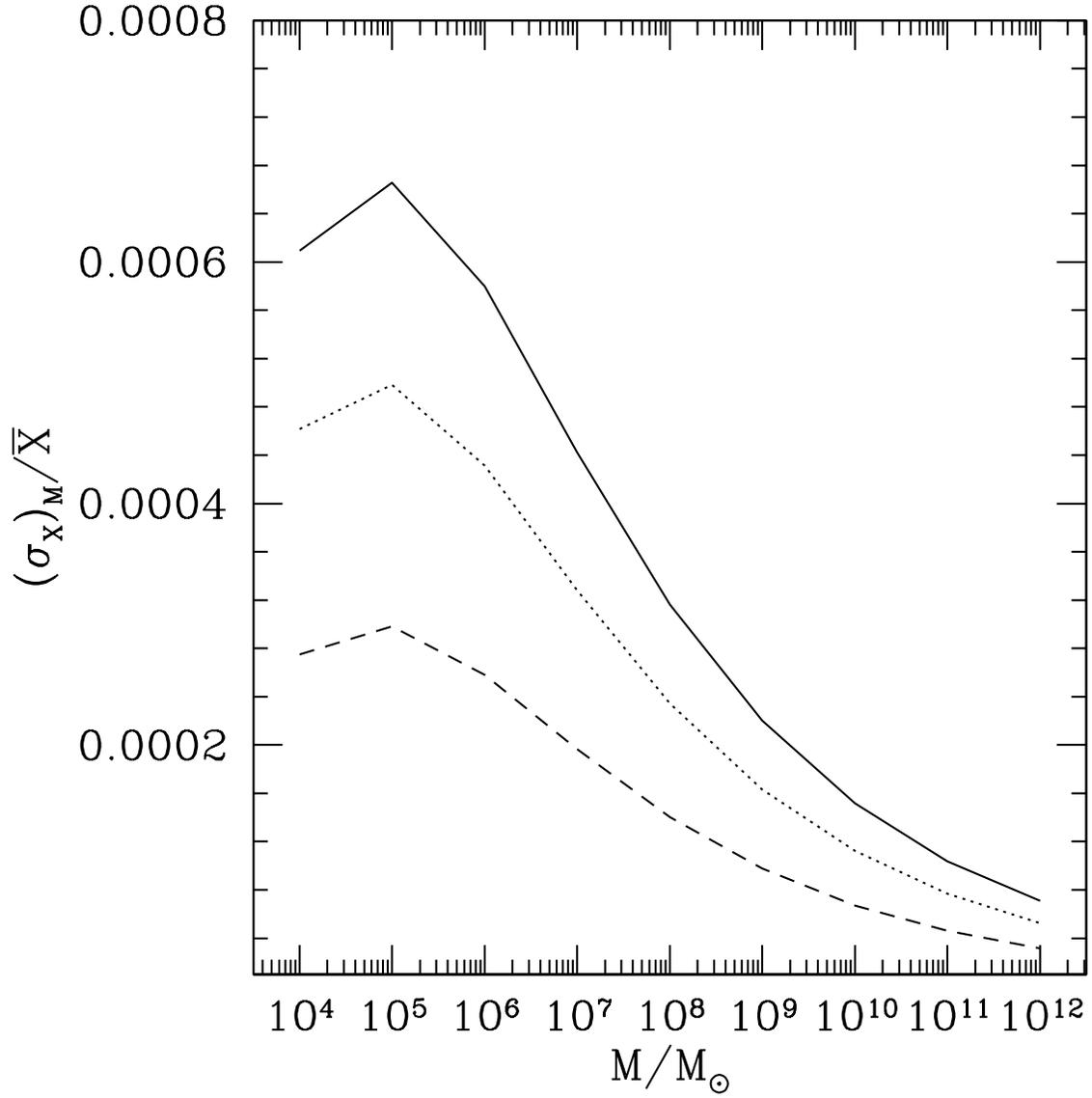} \figcaption[fig1]{ Standard deviations of abundances
for $z_{reion}=8$. The solid line is the standard deviation for $^4$He, the
dotted line is for $^7$Li, and the short dashed line is for D.
\label{fig:sigplot}}
\end{center}
\end{figure}

\begin{figure}
\begin{center}
\plotone{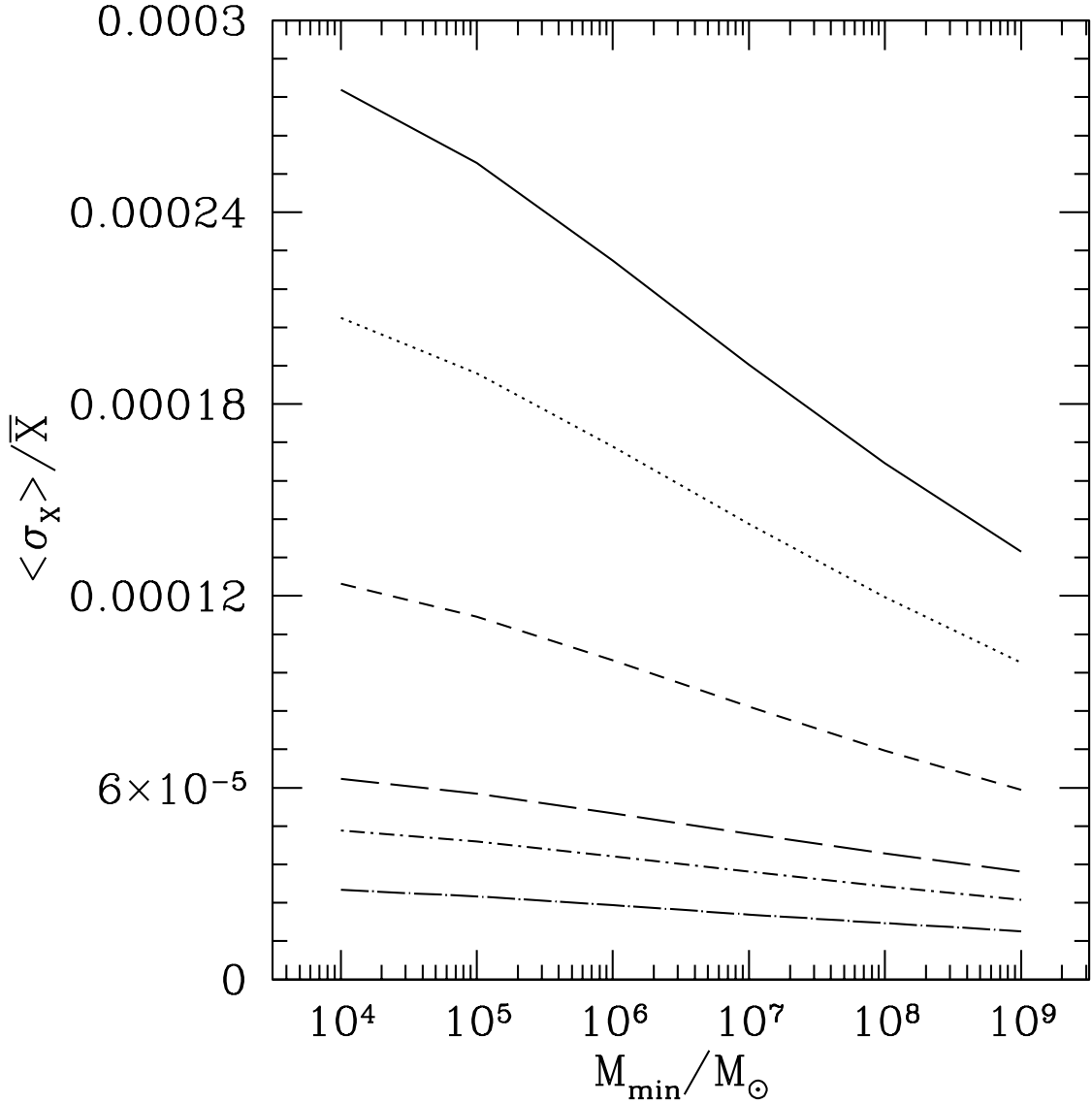} \figcaption[fig1]{ Standard deviations averaged
over all collapsed objects in the Universe as a function of the minimum
allowed mass scale. We show two cases for the collapse redshift: (i)
$z_{coll}=2.5$, for which the solid line is the standard deviation for
$^4$He, the dotted line is for $^7$Li, and the short-dashed line is for D;
(ii) $z_{coll}=0$ for which the long-dashed line refers to $^4$He, the
dot-short dashed line refers to $^7$Li, and the dot-long dashed line refers
to D.
\label{fig:siguniavg}}
\end{center}
\end{figure}

We carry out the integrations as with equation (\ref{eq:mdelint}) and our
results are plotted as a function of mass scale in
Figure~\ref{fig:sigplot}.  Comparing Figures~\ref{fig:meandev}
and~\ref{fig:sigplot}, we see that the relative importance of abundance
fluctuations depends on the mass of the object in question.  On small mass
scales, abundances differ from their primordial values by an approximately
uniform amount; but as the object size increases fluctuations from the mean
are more likely to produce a difference from the primordial values.  We
also average the standard deviation over all collapsed objects in the
Universe.  Our prescription for this is identical to that given by equation
(\ref{eq:uniavg}), with $\sigma_M$ taking the place of $Y_M$.  These
results are plotted versus the minimum mass scale for galaxies in
Figure~\ref{fig:siguniavg}.  Comparing this to Figure~\ref{fig:unimean}, we
again see that the relative importance of fluctuations increases with
increasing mass scale.

\section{Observational Data}
\label{sec:obs}
Though much recent work has focused on uncovering the primordial $^7$Li and
D abundances (e.g., Burles \& Tytler 1998a,b; Tytler et~al. 1999; Webb
et~al. 1997; Ryan et~al. 2000), the $^4$He abundance with a reported
uncertainty of about $1\%$, is still more accurately determined.  However,
even this accuracy is at least an order of magnitude coarser than the mean
effect of diffusion for low-mass objects.  But, as discussed in
\S~\ref{sec:calc} the $^4$He abundance is also expected to vary from point
to point within an object.  In this section, we re-examine the $^4$He
abundance data presented in Izotov \& Thuan (1998) to determine whether or
not such fluctuations are currently detectable.

In Izotov \& Thuan (1998), the helium abundance is assumed to be linear in
metallicity for low metallicity objects.  Each data point consists of a
measurement of the metallicity, $Z_i$, a measurement of the helium
abundance $Y_i$, and estimations for the uncertainties in each,
$\epsilon_i$ and $\delta_i$.  A straight line is fit to the data, the
intercept of which reveals the primordial (zero-metallicity) helium
abundance.  The model we use is slightly different.  Because of diffusion
we cannot directly associate $^4$He abundance measurements of
zero-metallicity objects with the primordial $^4$He abundance.  Not only do
we expect an object of a particular mass to have a $^4$He abundance
differing from the primordial value as shown in Figure~\ref{fig:meandev},
but also we expect that different measurements of the single object would
display an intrinsic scatter about this mean as shown in
Figure~\ref{fig:sigplot}.  Therefore, we fit the data to the following
model.  If an object has metallicity $Z$, the probability that it will have
a $^4$He mass fraction $Y$ is denoted $g\left( Y|Z\right) dY$.  We account
for measurement uncertainties by defining $h\left( Y_i | Y\right) $ to be
the probability of measuring $Y_i$ if the object actually has abundance
$Y$, and $f\left( Z | Z_i \right) $ to be the probability that an object
has metallicity $Z$ if $Z_i$ is what is observed.  Thus, the total
probability of a measurement simultaneously obtaining $Y_i$ and $Z_i$ is
\begin{equation}
P(Y_i | Z_i) =\int f\left( Z | Z_i \right)
g\left( Y | Z\right) h\left( Y_i | Y\right) dY dZ .
\label{eq:probgen}
\end{equation}
Both $Z$ and $Y$ 
are intrinsically positive, but we here assume that the 
uncertainties in the measurments are small enough so that the
probability distributions $f$ and $h$ may be taken to be Gaussians,
\begin{equation}
f\left(Z |Z_i \right) = \frac{1}{\sqrt{2\pi\epsilon_i^2}}
\exp\left[-\frac{\left( Z-Z_i \right)^2}{2\epsilon_i^2}\right] ,
\label{eq:zmodel}
\end{equation}
\begin{equation}
h\left( Y_i |Y \right) = \frac{1}{\sqrt{2\pi\delta_i^2}}
\exp\left[-\frac{\left( Y-Y_i \right)^2}{2\delta_i^2}\right] .
\label{eq:ymodel}
\end{equation}
As for our model, we assume that
\begin{equation}
g\left( Y | Z\right) = \frac{1}{\sqrt{2\pi\sigma^2}}
\exp\left[-\frac{\left( Y-mZ-b \right)^2}{2\sigma^2}\right] .
\label{eq:mcmodel}
\end{equation}
The parameters to be determined by the data are
$m$, which contains the physics of the metallicity dependence; $b$, 
which is the mean value for the $^4$He abundance;
and $\sigma$, which represents the intrinsic scatter about the mean.

\begin{figure}
\begin{center}
\plotone{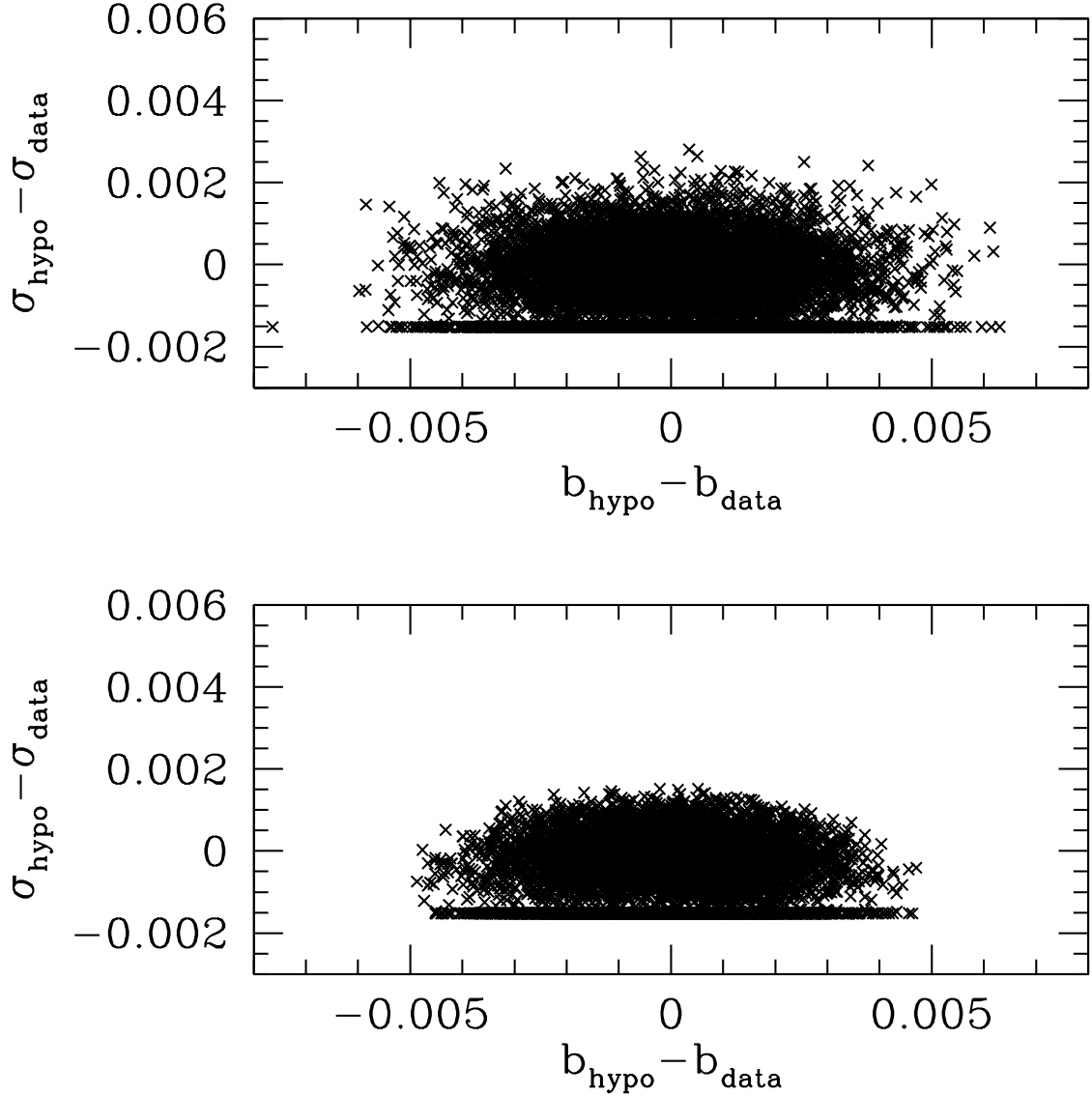} \figcaption[fig1]{Results from $10^4$ Monte Carlo data
sets based on the model described in \S 5.  The top panel shows all
hypothetical $\left(b_{hypo}, \sigma_{hypo}\right) $ pairs relative to the
observed values ($b_{data}$,$m_{data}$), while the bottom panel show the
$6,730$ pairs most likely to occur.  The extended horizontal concentration
of points near $-0.0015$ occurs because of the physical constraint that all
values of $\sigma$ be greater than zero.  This indicates that we have not
yet reached the level of accuracy required to put a lower bound on
$\sigma$.
\label{fig:mcplots}}
\end{center}
\end{figure}

We use the integration in equation (\ref{eq:probgen}) to obtain
\begin{equation}
P\left( Y_i | Z_i\right)=\frac{1}{\sqrt{2\pi}}\frac{1}{\sqrt{\sigma^2
+m^2\epsilon_i^2+\delta_i^2}}\exp \left[ -\frac{ 
\left( Y_i -mZ_i -b\right)^2}{2\left( \sigma^2 +
m^2\epsilon_i^2+\delta_i^2\right)} \right] .
\label{eq:probint}
\end{equation}
If we determine the metallicity and helium abundance for $N$ objects,
the probability for the entire measurement to occur is
\begin{equation}
P_N=\prod_i^N P\left( Y_i | Z_i\right) .
\label{eq:probtot}
\end{equation}
To maximize this probability with respect to the parameters $m$, $b$,
and $\sigma$, we minimize the $\chi^2$-like function 
\begin{equation}
\chi^2 = \sum_{i=1}^N \left[ \ln \left( \sigma^2 +\delta_i^2
+m^2\epsilon_i^2\right) + \frac{ \left( 
Y_i-mZ_i-b\right)^2}{\sigma^2 +m^2\epsilon_i^2
+\delta_i^2} \right] .
\label{eq:chitwo}
\end{equation}
A downhill simplex method and a direction set method yielded almost
identical parameters for the minimum $\chi^2$: $m=4.8\times 10^{-3}$,
$b=0.244$, and $\sigma=1.52\times 10^{-3}$.  We used a Monte Carlo method
to set confidence limits on these results. Because of random errors, the
observed data points $Z_i$ and $Y_i$ are not a unique realization of the
true values of the quantities.  Instead, for a given measurement of $Z_i$,
there are an infinite number of $Y_i$ that could have been measured.  The
probabilities for obtaining these hypothetical measurements are assumed to
be given by equation (\ref{eq:probint}).  We have generated $10^4$ of these
hypothetical data sets by substituting the observed values for $m$, $b$,
and $\sigma$ into equation (\ref{eq:probint}).  For each hypothetical data
set, we again minimized equation (\ref{eq:chitwo}) to obtain a new triplet
of parameters $m$, $b$, and $\sigma$.  Figure~\ref{fig:mcplots} shows the
resulting intercepts $b$ plotted against the scatters $\sigma$.  The
uncertainty in our best-fit sigma and intercept are large, and so the
observational situation must improve before we could settle on the true
parameters.

\section{Conclusions}
\label{sec:sandc}

We have explored the effect of diffusion during the linear growth of
density perturbations in the neutral IGM before the redshift of
reionization and have found that element abundances within collapsed
objects differ from their primordial values. Element diffusion leads to two
related effects.  First, the mean element abundances of a collapsed object
could differ from the primordial values; and second, diffusion could result
in abundance fluctuations within a single object.  For objects with a total
mass below $10^7$M$_{\odot}$ the former effect dominates; but for larger
objects the latter effect is more important.  The magnitude of the effect
is $\sim 0.005$--$0.2\%$, depending on the mass of the object and the
element.  We have then used a Press-Schechter mass function to average
these distributions over all collapsed objects in the Universe, and again
found the net effect to be $< 0.1\%$.  This effect is yet too small to be
detected since the best observations of element abundances currently have
accuracies of $\sim 1\%$.

If the effect of diffusion on element abundances will eventually be
measured with sufficient precision in low-metallicity systems, the results
could provide important clues about early structure formation in the IGM.
In particular, the mean deviations within objects are sensitive to the
minimum mass of a galaxy halo.  Also, the overall magnitude of diffusion is
sensitive to the redshift of reionization since diffusion is brought to a
halt at that time.  And as determinations of element abundances get more
and more accurate, diffusion will eventually have to be accounted for in
order to make comparisons with the predictions of big bang nucleosynthesis.

\acknowledgements

We thank Rennan Barkana for useful comments on the manuscript.  This work
was supported in part by NASA grants NAG 5-7039, 5-7768, and by NSF grants
AST-9900877, AST-0071019.


\newpage

\newpage

\newpage

\newpage

\newpage

\newpage

\newpage

\newpage

\newpage
\end{document}